\documentclass[aps,prl,twocolumn,preprintnumbers,10pt,showpacs]{revtex4-1}
\usepackage{amsmath,bm,amssymb}

\usepackage{graphicx}
\usepackage{psfrag}
\usepackage[caption=false]{subfig}

\newcommand\eps{\epsilon}

\usepackage{slashed}

\allowdisplaybreaks

\begin{document}

\preprint{MITP/15-101, ZU-TH 36/15}

\title{Analytic form of the two-loop planar five-gluon all-plus-helicity amplitude in QCD}

\author{T.\ Gehrmann$^a$, J.\ M.\ Henn$^b$, N.\ A.\ Lo Presti$^a$}

\affiliation{$^a$ Department of Physics, University of
  Z\"urich, Winterthurerstrasse 190, CH-8057 Z\"urich, Switzerland\\
$^b$ 
PRISMA Cluster of Excellence, 
Johannes Gutenberg University, 55099 Mainz, Germany}


\begin{abstract}
Virtual two-loop corrections to scattering amplitudes are a key ingredient to precision physics at collider experiments. 
We compute the full set of planar master integrals relevant to five-point functions in massless QCD, and 
use these to derive an analytical expression for the two-loop five-gluon  all-plus-helicity amplitude. 
After subtracting terms that are related to the universal infrared 
and ultraviolet
pole structure, we obtain a remarkably simple and 
compact 
finite
remainder function, consisting only of
dilogarithms. 
\end{abstract}

\maketitle

The precise theoretical description of scattering reactions of elementary particles relies on the 
perturbation theory expansion of the scattering amplitudes describing the process under consideration. In this expansion, 
higher perturbative orders correspond to more and more virtual particle loops. At present, one-loop corrections 
can be computed to scattering amplitudes of arbitrary multiplicity, while two-loop corrections are known only for 
selected two-to-one annihilation or two-to-two scattering processes. 

For many experimental observables at higher multiplicity, 
a substantial increase in statistical precision can be expected from the CERN LHC in the near future.
Perturbative predictions beyond one loop will be in demand for many precision applications of these data, for example 
in improved extractions of standard model parameters or in search for indirect signatures of new 
high-scale physics in precision observables. 

Progress on multiloop corrections to high-multiplicity amplitudes requires significant advances in two directions. 
Feynman-diagrammatic approaches to the computation of these amplitudes yield enormously large expressions that 
contain many thousands of different Feynman integrals. These integrals are related among each other 
through Poincar\'e invariance and symmetries, such that only a limited set of independent 
so-called master integrals will remain in the final answer for a scattering amplitude. To express a generic two-loop multiparton 
amplitude in terms of the relevant master integrals (ideally circumventing the large algebraic complexity at intermediate stages 
that is generated by working in terms of Feynman diagrams) is a yet outstanding problem. 
A particular example where the reduction to a basis set of integrals was achieved~\cite{badger5g,badger5gfull} 
is the two-loop five-gluon 
helicity amplitude with all helicities positive. In this case, the application of on-shell techniques led to 
a particularly compact integrand, which motivated a specific choice of basis integrals (which do not necessarily form a minimal 
set in the sense of being master integrals). In~\cite{badger5g}, these integrals were 
evaluated numerically for selected kinematical points. 
Although this specific helicity amplitude is not contributing to the second-order correction to the three-jet cross section (due to its vanishing at tree level), it provides an ideal testing laboratory for new calculational concepts and methods that will carry over to the general helicity case, 
as previously in the case for the four-point two-loop amplitudes \cite{Bern:2000dn}.

The other major challenge in the calculation of multileg multiloop amplitudes lies in the evaluation of the master integrals. While the full 
set of Feynman integrals at one loop is known for all configurations of internal masses and external kinematics, only 
specific integrals at low multiplicity (typically four external legs; see, however, \cite{Drummond:2010cz,Dixon:2011nj}) are known at two loops and beyond. 
In principle, these integrals can be evaluated using purely numerical methods, as for example iterated sector 
decomposition~\cite{secdec,fiesta}. In practice, these methods turn out to be too slow to allow for an 
efficient evaluation of multiloop integrals when sampling the multidimensional phase space, as required for the evaluation of scattering cross sections.  

On the other hand, analytical expressions for the integrals allow to uncover universal structures in the amplitudes and enable the 
study of limiting kinematical behaviour, thereby advancing our understanding of the high-order structure of perturbative quantum field theory. 
The analytical understanding of the most 
basic amplitudes at a given multiplicity and loop order  
is moreover an important catalyst enabling further progress towards more complicated processes at the same 
multiplicity. Important examples are the reconstruction of amplitudes from constraints obtained in specific limits \cite{Bern:1994zx}, the further development and validation of integrand reduction techniques, and methods for finding appropriate integral bases \cite{ArkaniHamed:2010gh,henn}.

In this letter, we make use of advances in techniques for analytically evaluating loop integrals \cite{henn},
and we compute for the first time the full set of planar master integrals relevant to massless two-loop five-point scattering. 
We apply these integrals to the five-gluon all-plus helicity amplitude. 
This is the first analytic result of a genuine $2 \to 3$ two-loop amplitude in QCD.
We find that after subtraction of the universal infrared singular terms, our analytic formula is remarkably simple.
It can be written in terms of dilogarithms, with prefactors that are well-behaved in collinear limits.
This simplicity, which is reminiscent of results in ${\mathcal N}=4$ super Yang-Mills (SYM), may help to uncover new structural properties 
of multileg multiloop amplitudes and lead to considerable simplifications in their calculation.

\section{Master integrals for two-loop five-point functions}

\begin{figure}[t]
 \centerline{
  \hspace{-0.3cm}\subfloat[]{\includegraphics[scale=0.20]{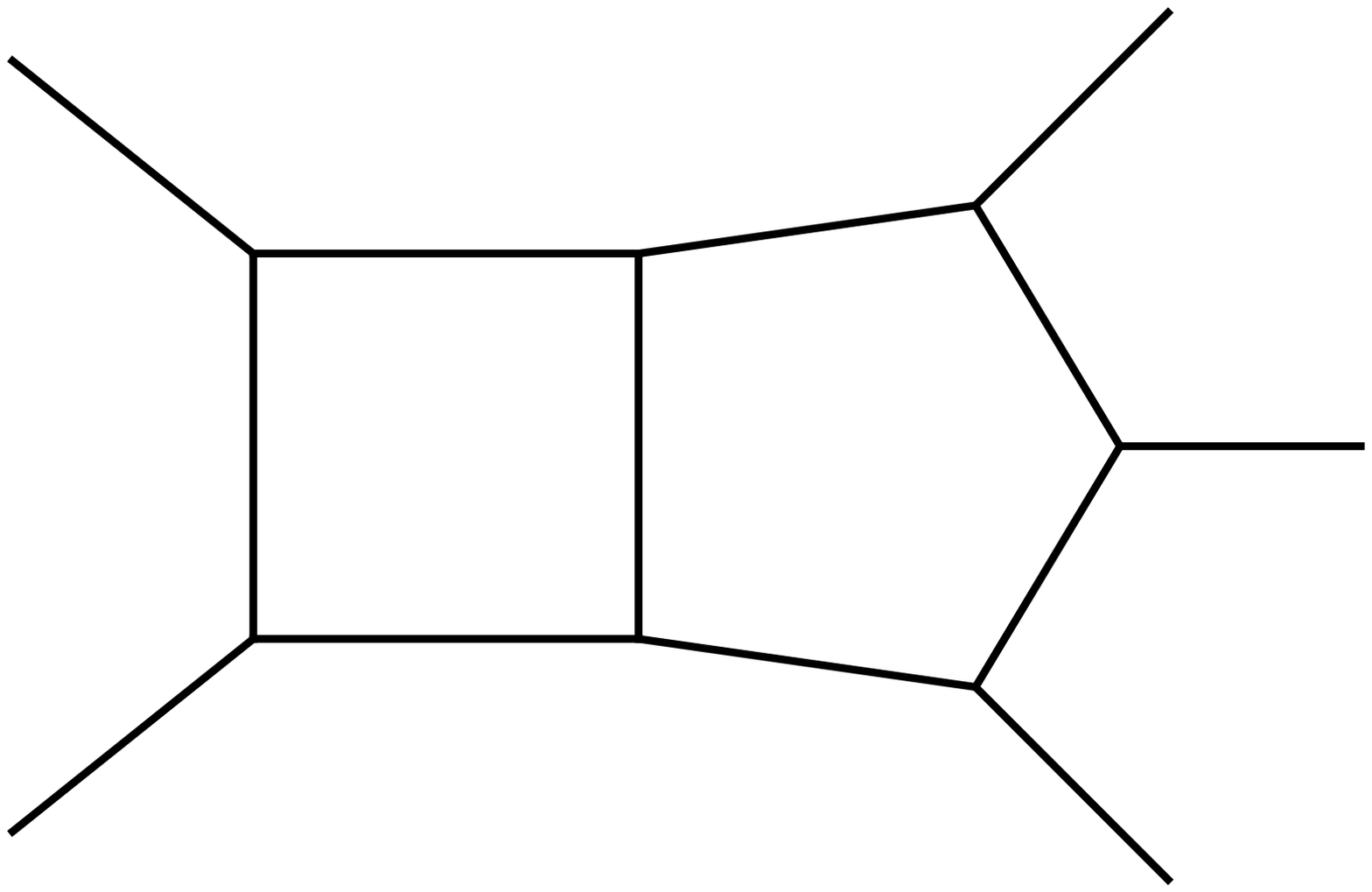}}\quad\quad
   \subfloat[]{\includegraphics[scale=0.20]{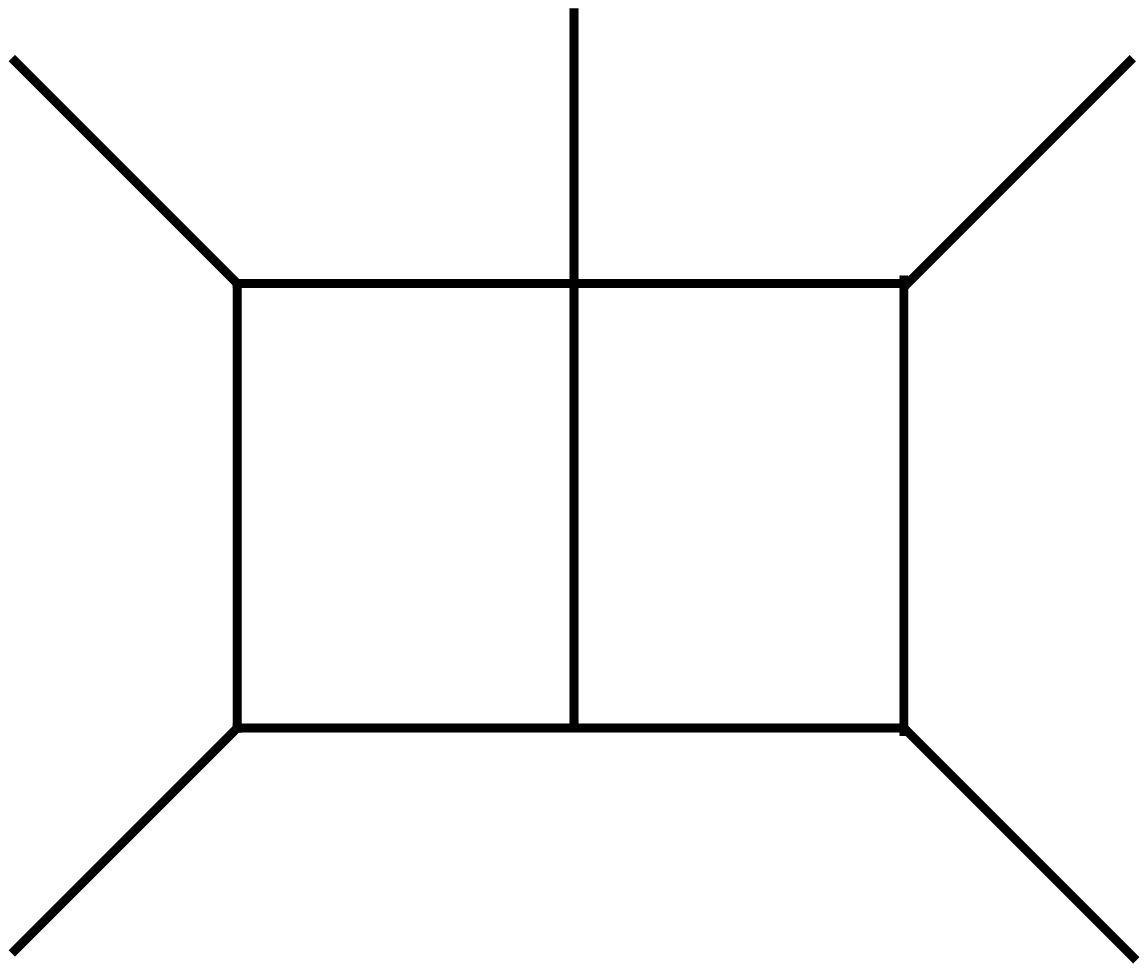}}
  }
  \vspace{-0.2cm}
 \centerline{
        \subfloat[]{\includegraphics[scale=0.20]{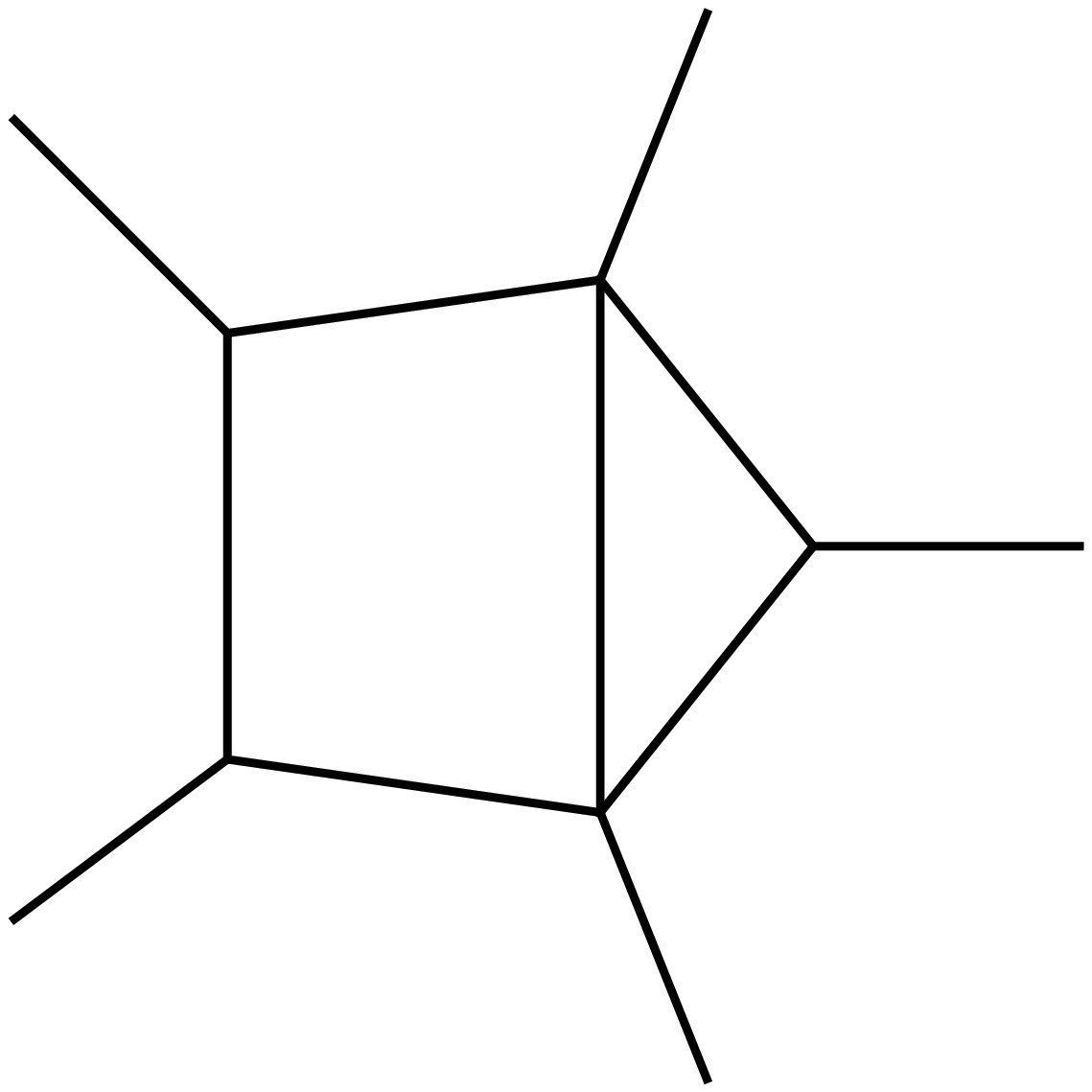}}\quad\quad
        \hspace{0.4cm}
    \subfloat[]{\includegraphics[scale=0.20]{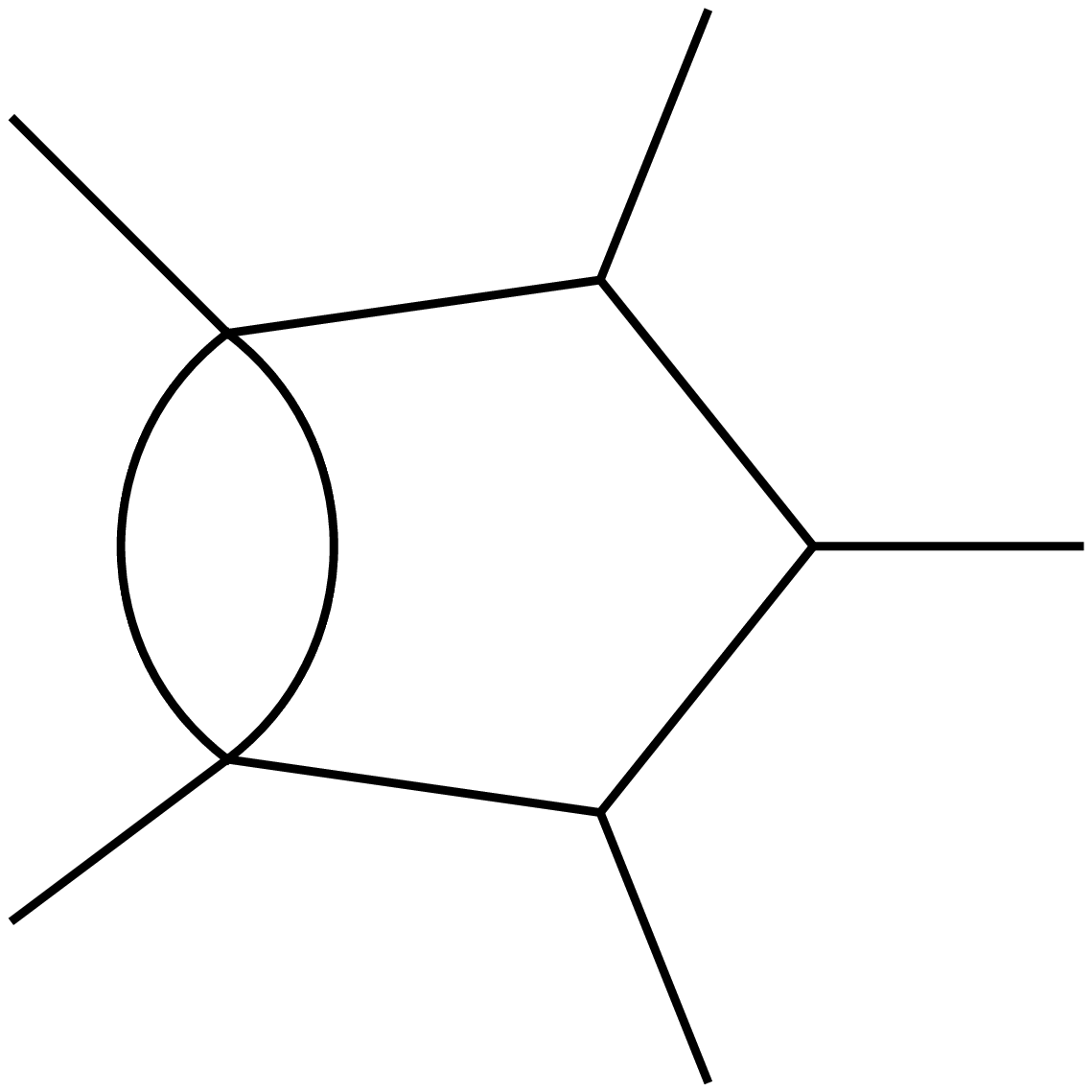}}
 }
  \caption{Genuine five-point planar two-loop integrals.}
 \label{fig:pentagonintegrals}
\end{figure}

Feynman integrals in dimensional regularization in $4-2\eps$ 
dimensions are invariant under Poincar\'e transformations. By 
applying these transformations at the integrand-level, one obtains 
nontrivial linear relations among 
different integrals, the integration-by-parts \cite{chet}
relations.
These relations can be used to reduce the large number of Feynman integrals relevant to a particular process 
to a much smaller number of so-called master integrals. 
This  reduction is typically carried out using a lexicographic ordering of the 
integrals~\cite{laporta}, implemented in computer algebra routines, for example in the codes 
\cite{fire} or
\cite{reduze}. 

The type of master integrals that appear in a given process
depends only on the external kinematics, and on possible internal propagator masses. 
All two-loop five-parton amplitudes relate to a common set of master integrals: 
massless on-shell five-point functions at two loops. 
These can be further classified into genuine five-point functions, four-point functions with one off-shell leg, 
three-point functions with up to two off-shell legs and off-shell two-point functions. 
Up to the four-point level, these functions appeared in the context of the derivation of the two-loop amplitude 
for $\gamma^\star \to 3$~jets~\cite{3jamp} and were already computed long ago~\cite{gr,3jint}. 
The genuine five-point functions depend on five independent Mandelstam invariants, 
\begin{displaymath}
v_1 = s_{12},\quad 
v_2 = s_{23},\quad 
v_3 = s_{34},\quad 
v_4 = s_{45},\quad 
v_5 = s_{51}\,,
\end{displaymath}
where $s_{ij}=2 p_{i} \cdot p_{j}$.
They are therefore considerably more complicated than the four-point functions, since the latter depend on three variables only.
We find in total 25 new integrals (10 planar and 15 nonplanar). 
The planar integrals can be given in terms of four integral topologies, displayed in Figure~\ref{fig:pentagonintegrals}. 
There are $3,3,2$, and $2$ master integrals for topologies (a), (b), (c), and (d), respectively.

To compute the integrals, we derive differential equations for them in the $v_{i}$.
The system of differential equations 
is then brought into a canonical form~\cite{henn} by means of 
a transformation of the basis of master integrals to integrals having unit leading singularities \cite{ArkaniHamed:2010gh}. 
The canonical form we find is
\begin{align}\label{decanonical}
d \vec{f}(v_i; \epsilon) = \eps \left[ \sum_{i} a_{i} d\log(\alpha_{i}) \right] \vec{f}(v_i; \epsilon) \,,
\end{align}
where $\vec{f}$ is the set of $61$ master integrals, the differential $d$ comprises partial derivatives w.r.t. $v_{i}$,
and $a_{i}$ are constant (kinematic and $\eps$-independent) matrices. 
The collection of letters $\alpha_{i}$ specify the function alphabet $\mathbb{A}$. The latter is given by
\begin{align}\label{functionalphabet}
& \left\{ v_1 , v_3 + v_4, v_1 -  v_4, v_1+v_2 - v_4,  \Delta, \frac{ a - \sqrt{\Delta}}{a+ \sqrt{\Delta}}  \right\}  \,,
\end{align}
and cyclic permutations thereof. 
Here, 
\begin{align}
a =& v_1 v_2 - v_2 v_3 + v_3 v_4 -v_1 v_5 - v_4 v_5
\,,
\end{align}
and 
the Gram determinant 
$\Delta = | 2 p_{i}\cdot p_{j} |$, with $1 \le i,j \le 4$. 
It is interesting to note that 
$a ={\rm tr}[\slashed{p}_{4}\slashed{p}_{5}\slashed{p}_{1}\slashed{p}_{2}]$
and
$\Delta = ({\rm tr}_{5})^2$, where ${\rm tr}_{5} = {\rm tr}[ \gamma_{5} \slashed{p}_{4}\slashed{p}_{5}\slashed{p}_{1}\slashed{p}_{2}]$.

The full set of master integrals can be obtained by direct integration of the differential equations, 
order-by-order in $\eps$, in terms of Chen iterated integrals \cite{chen}.
For a practical application of the latter to multivariable Feynman integrals, including their numerical evaluation, see \cite{Caron-Huot:2014lda}.
The boundary conditions are determined from consistency conditions, such as the absence of unphysical branch cuts.
This is particularly simple to implement in the canonical form (\ref{decanonical}) of the differential equations, cf. \cite{Henn:2013nsa}. 

Massless scattering is naturally parametrized using momentum twistor variables \cite{Hodges:2009hk} that solve both the on-shell as well as the momentum conservation constraints. 
We find that in these variables  the alphabet (\ref{functionalphabet}) becomes rational. This implies that, when expressed in terms of these variables, the Chen iterated integrals degenerate to
multiple polylogarithms~\cite{hpl,ghpl}, for which efficient and precise numerical representations 
exist~\cite{vollinga}. All subtopologies at four points and below are recomputed in 
terms of the momentum twistor variables, yielding agreement with earlier results~\cite{gr,3jint}. 

We further validate our integrals by analytically computing all five-gluon amplitudes in $\mathcal{N}=4$ super Yang-Mills at two loops. Their expression was initially conjectured in \cite{Anastasiou:2003kj}, tested numerically in \cite{neq4sym5ptnum}, and proven in \cite{Drummond:2007au} from a Ward identity for 
dual conformal 
symmetry. Our calculation is the first
direct analytical one, and we find complete agreement with the above references.

Technical details on the determination of the five-point master integrals will be documented in a 
separate publication~\cite{upcoming}.

\section{Result for all-plus amplitude}

We consider the unrenormalized all-plus five-gluon amplitude at leading color:
\begin{align}
{\mathcal A}_{5}(1^+ 2^+ 3^+ 4^+ 5^+) |_{\rm leading \; color}=& g^3  \sum_{L \ge 1} \left( g^2 N c_{\Gamma}\right)^L  \nonumber\\
&\hspace{-4cm} \times \sum_{\sigma \in S_{5}/Z_{5}} {\rm tr}(T^{a_{\sigma(1)}} T^{a_{\sigma(2)}} T^{a_{\sigma(3)}} T^{a_{\sigma(4)}} T^{a_{\sigma(5)}}) \nonumber\\
&\hspace{-4cm} \times A^{(L)}_{5}(\sigma(1)^{+}  \sigma(2)^{+} \sigma(3)^{+} \sigma(4)^{+} \sigma(5)^{+})\,.
\end{align}
Here $S_{5}/Z_{5}$ denote all noncyclic rotations of five points,
and \cite{Bern:2000dn}
\begin{align}\label{cgamma}
c_{\Gamma}
= \frac{1}{(4\pi)^{2-\eps}}
\frac{\Gamma(1+\eps)\Gamma^2(1-\eps)}{\Gamma(1-2\eps)}.
\end{align}
Since the amplitude vanishes at tree level, it is finite at the one-loop level \cite{Bern:1993mq},
\begin{align}\label{oneloop}
A^{(1)}_{5}
=  R\; F_{5}^{(1)}+ {\mathcal O}(\eps)\,,
\end{align}
with 
$R = i/6/({\langle 1 2 \rangle \langle 2 3 \rangle \langle 3 4 \rangle \langle 4 5  \rangle \langle  5 1  \rangle})$
and
\begin{align}
F^{(1)}_{5} =&\; v_1 v_2 + v_2 v_3 + v_3 v_4 + v_4 v_5 +v_5 v_1 + {\rm tr}_{5} \,.
\end{align}
At two loops, the infrared and ultraviolet divergent terms can be predicted in terms of the one-loop result. 
This motivates the definition of a finite remainder $F^{(2)}_{5}$ according to
\footnote{The absence of an explicit $1/\eps$ pole in Eq. (\ref{definitionremainder}) 
is due to an
exact cancellation of ultraviolet and infrared terms in unrenormalized two-loop amplitudes that are
finite at one loop. After UV renormalization, the well-known infrared pole structure
\cite{Catani:1998bh} is recovered.}
 \begin{align}\label{definitionremainder}
A^{(2)}_{5} =& A^{(1)}_{5} \left[ - \sum_{i=1}^{5} \frac{1}{\epsilon^2} \left(\frac{\mu^2}{-v_{i}}\right)^{\epsilon} 
\right] + R \;F^{(2)}_{5}  + {\mathcal O}(\eps)\,.
\end{align}
We use the integral representation of \cite{badger5g} and express it in terms of our basis of integrals.
Plugging in the solution for the $\eps$-expansion of the latter,
we analytically verify the divergence structure of
Eq. (\ref{definitionremainder}). To define the finite remainder function, the expansion of (\ref{oneloop}) to order 
$\epsilon^2$ is derived, which involves the one-loop massless pentagon integral to this order,
computed from its differential equation. 
In the finite remainder, remarkably all Chen iterated integrals of weight one, three and four cancel out.
We then express the remaining weight-two functions in terms of dilogarithms, and find the following expression for the finite remainder: 
\begin{align}\label{resultF5nice}
F^{(2)}_{5} =& \frac{5 \pi^2}{12} F_{5}^{(1)} 
+  \sum_{i=0}^{4}  \sigma^{i} \left\{ \frac{v_5 
 {\rm tr}\left[ (1-\gamma_{5})   \slashed{p}_{4}  \slashed{p}_{5}  \slashed{p}_{1}  \slashed{p}_{2}  \right]  
  }{ (v_2 + v_3 -v_5) }
  I_{23,5}
  \right. \nonumber \\
 &\hspace{-0.7 cm} \left. + \frac{1}{6} \frac{ {\rm tr}\left[ (1+\gamma_{5})   \slashed{p}_{4}  \slashed{p}_{5}  \slashed{p}_{1}  \slashed{p}_{2}   \right]^2 }{v_1 v_4 } + \frac{10}{3} v_1 v_2 +\frac{2}{3} v_1 v_3  \right\} \,. 
 \end{align}
where $\sigma^{i}$ cyclically shifts all indices (of $p$, $v$, and $I$) by $i$, and where
%\begin{align}
%F_{23,5} =& \frac{1}{2} {\rm Li}_{2}\left(1-\frac{v_5}{v_2}\right) - \frac{1}{2} {\rm Li}_{2}\left(1-\frac{v_{2}}{v_{5}}\right) +  \frac{1}{2} {\rm Li}_{2}\left(1-\frac{v_5}{v_3}\right) - \frac{1}{2} {\rm Li}_{2}\left(1-\frac{v_{3}}{v_{5}} \right) \nonumber \\
%& + \frac{1}{4} \log^2 \frac{v_2}{v_{5}} + \frac{1}{4} \log^2 \frac{v_3}{v_{5}} - \log \frac{v_2}{v_5} \log \frac{v_3}{v_5} + \zeta_2 \,.
%\end{align}
\begin{align} \label{formulaonemassbox}
I_{23,5} =& \zeta_2 +   {\rm Li}_{2}\left[\frac{(v_5-v_2)(v_5-v_3)}{v_2 v_3 }\right] \nonumber \\
& -   {\rm Li}_{2}\left[\frac{v_5-v_3}{v_2}\right]
-   {\rm Li}_{2}\left[\frac{v_5-v_2}{v_3}\right] \,.
\end{align}
Note that Eq. (\ref{resultF5nice}) contains both parity odd and even terms.
We remark that the trace can also be written in a natural way using momentum twistors.

We compared our analytical result for the unrenormalized two-loop amplitude (\ref{definitionremainder}) with the numerical 
values quoted in~\cite{badger5g} for specific phase space points in the Euclidean region, finding full agreement. 
In the Euclidean region, this expression is single-valued and real. 
We note that Eq. (\ref{formulaonemassbox}) can be rewritten in a form where this is manifest,
and that our result can straightforwardly be analytically continued to other kinematical regions.
 
The result above is for pure Yang-Mills theory. 
We would like to mention that the full $n_{f}$ dependence can be reconstructed in a 
simple way: the $n_{f}^2$ terms only come from a restricted class of diagrams, and the remaining $n_{f}$ terms 
are fixed by supersymmetry \cite{Bern:2002tk}.

\section{Limits}

\begin{figure}
 \centerline{
  \includegraphics[scale=0.5]{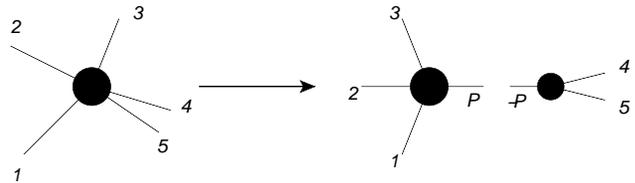}
  }
  \caption{Five-particle amplitude factorizing into four-point amplitudes and splitting functions in the collinear limit.}
 \label{fig:splitting}
\end{figure}

Scattering amplitudes have universal factorization properties in soft and collinear limits.
They serve as an important check of our result.

We take the $p_{4} || p_{5}$ collinear limit, without loss of generality.
In the limit, one expects (cf. Fig.~\ref{fig:splitting})
\begin{align}
A_{5}^{(2)}(1^{+}, 2^{+}, 3^{+}, 4^{+}, 5^{+}) \stackrel{p_{4} || p_{5}}{\longrightarrow} \label{collinear}  &\\ \nonumber 
& \hspace{-4cm} \phantom{+}\,\,\, A_{4}^{(1)}(1^{+},2^{+},3^{+},P^{+}) \, {\rm Split}^{P \to 45\, (1)}(-P^{-},4^{+},5^{+})
\\
& \hspace{-4cm}+   A_{4}^{(1)}(1^{+},2^{+},3^{+},P^{-})  \, {\rm Split}^{P \to 45\, (1)}(-P^{+},4^{+},5^{+})  \nonumber \\
& \hspace{-4cm} +  A_{4}^{(2)}(1^{+},2^{+},3^{+},P^{+}) \, {\rm Split}^{P \to 45\, (0)}(-P^{-},4^{+},5^{+}) \,. \nonumber
\end{align}
where `${\rm Split}$' are splitting amplitudes~\cite{splitloop}. 
The amplitudes appearing on the right hand side of Eq. (\ref{collinear}) can be found in \cite{Bern:2002tk}. 

Taking the collinear limit of  (\ref{definitionremainder}), we recover the structure predicted by (\ref{collinear}).
 It is interesting to note in this context that the second line of Eq. (\ref{resultF5nice}) contains 
 terms that behave as $[45]/\langle 4 5\rangle$ in this limit 
 (and are amplified by $1/\langle45\rangle$ from the overall factor $R$). 
 The latter reproduces a contribution from the helicity-violating one-loop splitting 
 function ${\rm Split}^{P \to 45\, (1)}(-P^{+},4^{+},5^{+})$. 

\section{Discussion and Outlook}

The simplicity of our result for the all-plus amplitude in QCD is reminiscent of similar results for six-gluon amplitudes in ${\mathcal N}=4$ SYM \cite{Goncharov:2010jf,Dixon:2011nj}.
In the latter case, the function alphabet is related to cluster algebras \cite{Golden:2013xva},
and it would be interesting to know whether this is also true for our five-point function alphabet of Eq. (\ref{functionalphabet}), or perhaps for a subset relevant for calculations up to finite parts of amplitudes.

It is interesting to investigate possible positivity properties \cite{Arkani-Hamed:2013jha,Arkani-Hamed:2014dca} of our result.
This is naturally done using momentum twistors. In QCD, the kinematics depends on the twistors $Z_{i}$ ($i=1,\ldots 5$),
and on an infinity twistor $Y$. It is tempting to treat the latter in the same way as the loop integration variable in the discussion of one-loop maximally 
helicity violating amplitudes in \cite{Arkani-Hamed:2014dca}.
This defines a kinematical region where in particular $v_{i}>0$ and ${\rm tr}[ (1-\gamma_{5}) \slashed{p}_{4} \slashed{p}_{5} \slashed{p}_{1} \slashed{p}_{2} ]<0$ (and cyclic). Interestingly, in this region, $F_{5}^{(1)}$ is positive (and the same holds for its $n$-point generalization \cite{Bern:1993qk}).
At two loops, we find that all terms are positive, except for the terms involving $I_{23,5}$, which are negative.
This can be seen by noting that $I_{23,5}/(v_2+v_3-v_5)$ is a one-loop one-mass box function in six dimensions.
It could be that the positivity properties at two loops are obscured by the infrared subtraction; see Eq. (\ref{definitionremainder}). 
However, one may speculate that the natural building blocks of definite sign that we found above point towards additional structure that is yet to be uncovered.

Our result for the planar master integrals provides the maximal set of polylogarithmic functions that can appear
in a generic planar massless five-particle scattering amplitude at two loops. 
Therefore we reduce the calculation of any such scattering amplitude to the determination of the algebraic coefficients 
accompanying the integral basis. This can be envisioned using a variety of related methods \cite{badger5g,Ossola:2006us}. In 
the one-loop case, the knowledge of the integral basis triggered a revolution in our ability to compute amplitudes, and we expect the same to occur here.

We would also like to mention that our result for the functional basis provides the foundation for bootstrap techniques to be used,
where one makes an ansatz for the remainder function of the type $\sum_{i} c_{i} F_{i}$, where $F_{i}$ are members of the functional basis, and $c_{i}$ are certain kinematic-dependent factors. See \cite{Dixon:2011pw,Dixon:2015iva} for applications of such a bootstrap approach to six-point 
amplitudes in $\mathcal{N}=4$ SYM. In the latter theory, the kinematic dependence of the $c_{i}$ is related, 
at least conjecturally, to leading singularities \cite{ArkaniHamed:2012nw}, 
simplifying the above ansatz. 
We anticipate that the basis of pentagon functions that we provide will be useful for
bootstrapping generic five-point amplitudes, especially when combined with an improved understanding 
of the possible algebraic coefficients that are allowed to appear in QCD.

An obvious next step consists in calculating the loop integrals in the 
nonplanar case (the method for finding an appropriate integral basis \cite{henn} does not rely on planarity) and applying them to the all-plus loop integrand that was found very recently \cite{badger5gfull}. 

Further generalizations include making one of the external legs off-shell,
where we anticipate Chen iterated integrals to play an even more important role 
than in the present work.
This will allow to describe processes such as the production of Higgs boson plus jets at NNLO.

\section{Acknowledgments}

We would like to thank Simon Badger for numerous discussions and for providing the numerical 
routines of~\cite{badger5g}, which were used to validate our results.
The expression (\ref{resultF5nice}) for the finite remainder agrees with the very
 recent result of~\cite{Dunbar:2016aux}, which uncovered a typo in the original version of our manuscript.
This work was supported in part by the Schweizer Nationalfonds under grant
200020-162487, as well as by the European Commission through
the ERC Advanced Grant ``MC@NNLO" (340983). 
J.M.H. is supported in part by a GFK fellowship and by
the PRISMA cluster of excellence at Mainz university.

\end{document}